\documentclass[aps,prl,twocolumn,superscriptaddress]{revtex4-1}
\usepackage{epsfig}
\usepackage{amssymb}
\usepackage{graphicx}
\usepackage{amsmath}
\usepackage{natbib}

\begin{document}

\title{Two-color spin noise spectroscopy: Using spin fluctuation correlations to reveal homogeneous linewidths within quantum dot ensembles}

\author{Luyi Yang}
\affiliation{National High Magnetic Field Laboratory, Los Alamos, NM 87545, USA.}
\author{P. Glasenapp}
\affiliation{Experimentelle Physik 2, Technische Universit\"{a}t Dortmund, D-44227 Dortmund, Germany.}
\author{A. Greilich}
\affiliation{Experimentelle Physik 2, Technische Universit\"{a}t Dortmund, D-44227 Dortmund, Germany.}
\author{D. R. Yakovlev}
\affiliation{Experimentelle Physik 2, Technische Universit\"{a}t Dortmund, D-44227 Dortmund, Germany.}
\affiliation{Ioffe Physical-Technical Institute, Russian Academy of Sciences, 194021 St. Petersburg, Russia.}
\author{M. Bayer}
\affiliation{Experimentelle Physik 2, Technische Universit\"{a}t Dortmund, D-44227 Dortmund, Germany.}\author{S.A. Crooker}
\affiliation{National High Magnetic Field Laboratory, Los Alamos, NM 87545, USA.}


\begin{abstract}
``Spin noise spectroscopy" (SNS) is a powerful optical technique for
probing electron and hole spin dynamics that is based on detecting
their intrinsic and random fluctuations while in thermal
equilibrium, an approach guaranteed by the fluctuation-dissipation
theorem. Because SNS measures fluctuation properties rather than
conventional response functions, we show that fluctuation
\emph{correlations} can be exploited in multi-probe noise studies to
reveal information that in general cannot be accessed by
conventional linear optical spectroscopy, such as the underlying
homogeneous linewidths of individual constituents within
inhomogeneously-broadened systems. This is demonstrated in an
ensemble of singly-charged (In,Ga)As quantum dots using two weak
probe lasers: When the two lasers have the same wavelength, they are
sensitive to the same QDs in the ensemble and their spin fluctuation
signals are correlated. In contrast, two probe lasers that are
widely detuned from each other measure different subsets of QDs,
leading to uncorrelated fluctuations. Measuring the noise
correlation versus laser detuning directly reveals the QD homogeneous
linewidth even in the presence of a strong inhomogeneous broadening.
Such noise-based correlation techniques are not limited to
semiconductor spin systems, but can be widely applied to any system
in which intrinsic fluctuations are measurable.
\end{abstract}


\maketitle

Inhomogeneous broadening is ubiquitous in the physical, chemical and
materials sciences, occurring whenever a collection of
nominally-equivalent constituents differ in, \emph{e.g.}, size,
shape, composition, conformation and/or local environment.  Notable
examples in nanoscale materials include ensembles of nanocrystals,
quantum dots, nanotubes, or molecules \cite{Alivisatos, Scholes, Bawendi}. In optically active systems,
such broadening typically leads to a spread of the constituents'
fundamental absorption or emission energies over an inhomogeneously
broadened band of spectral width $\gamma_{inh}$, which can be orders
of magnitude larger than the underlying \emph{homogeneous} linewidth
$\gamma_h$ of the individual constituents themselves \cite{Bawendi, Gammon,
Atature, Gerardot, WarburtonReview, Vamivakas}. Usually, however,
$\gamma_h$ is the essential quantity of interest, since $\gamma_h$
directly reveals (or at least constrains) the fundamental relaxation
rates and coherence times of the system, which are the crucial
parameters for many applications.

Unfortunately, $\gamma_h$ is generally inaccessible in
inhomogenously-broadened ensembles using conventional
low-power/linear optical spectroscopic techniques, which typically
measure time- and ensemble-averaged response functions (such as
absorption, photoluminescence, polarization, \emph{etc}). To
circumvent this limitation, various \emph{non-linear} optical
methods have been very successfully developed over the years to
extract $\gamma_h$ from inhomogeneously broadened ensembles;
examples include spectral-hole burning \cite{Berry, Palinginis} and
four-wave mixing methods \cite{Remillard, Borri, BorriReview,
Moody, MoodyTrion}, which necessarily rely on the excitation
and nonlinear optical response of the material. In parallel, a
multitude of optical techniques for isolating and measuring
\emph{single} particles \cite{Bawendi, Gammon, Atature, Gerardot,
WarburtonReview, Vamivakas} have also been developed to get around
the problem of inhomogeneous broadening in ensemble studies.

In this work we develop a novel low-power optical technique -- two-color spin
noise spectroscopy -- and demonstrate that it can be used to reveal
the underlying homogeneous linewidth $\gamma_h$ of the individual
constituents that make up an otherwise  inhomogeneously-broadened
ensemble. Specifically we apply these methods to reveal $\gamma_h$
of singly-charged (In,Ga)As quantum dots (QDs) within an
inhomogeously-broadened QD ensemble. Importantly, this optical technique operates in the linear/low-power regime and does not rely on any
excitation or nonlinear response of the system. The key point is that this technique is \emph{not} based
on conventional time-averaged response functions, but rather is
based on the \emph{intrinsic and random fluctuation properties} of
the system -- in this case, spin fluctuations. In particular, it
exploits the fact that spin fluctuations from different QDs in the
ensemble are uncorrelated in time. By measuring the degree of
correlation between two independent noise probes (two probe lasers
detuned from each other), we reveal the underlying homogeneous
absorption linewidth, $\gamma_h$, of positively-charged QDs in an
ensemble measurement -- information that is generally inaccessible
to conventional linear spectroscopy.

\begin{figure*}
\centering
\includegraphics[width=0.65\textwidth]{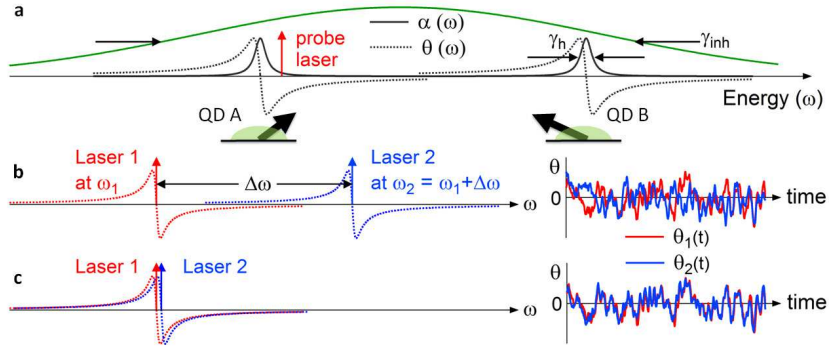}
\caption{\textbf{Correlated and uncorrelated spin fluctuations in
inhomogeneously broadened QD ensembles.} \textbf{(a)}, Illustrations depicting the inhomogeneously-broadened absorption band of a QD ensemble (green line), along with the homogeneously-broadened absorption and associated Faraday rotation spectra [$\alpha(\omega)$ and $\theta(\omega)$] of two representative singly-charged QDs in the ensemble (QDA and QDB). The probe laser that is shown is primarily sensitive to fluctuations of the spin in QDA, but not QDB. \textbf{(b)}, Cartoon showing two probe lasers at photon energies $\omega_1$ and $\omega_2$. The Faraday rotation (FR) sensitivity of each laser to spin fluctuations in QDs at energy $\omega$ is shown by the dotted lines (note that the probe lasers are not sensitive to spin noise from QDs exactly on resonance, because $\theta(\omega)$ is an odd function). Here, the two lasers are well-separated in energy ($\Delta \omega \gg \gamma_h$), and therefore they are sensitive to different subsets of QDs, so that the FR noise on the two lasers [$\theta_1(t)$ and $\theta_2(t)$] are largely \emph{uncorrelated} in time. \textbf{(c)}, The same, but for the case of small laser detuning ($\Delta\omega \leq \gamma_h$). Here, the two lasers probe predominantly the same dots and FR noise is \emph{correlated}, giving larger measured noise power $\langle[\theta_1(t)
+\theta_2(t)]^2 \rangle$.}\label{fig:1}
\end{figure*}

Optical spin noise spectroscopy (SNS) is a powerful and relatively
new technique for probing the dynamics of electron and/or hole
spins, that is based on measuring their intrinsic fluctuations while
they remain unperturbed and in thermal equilibrium \cite{CrookerGas,
ZapasskiiR, MullerPhysicaE}. This approach, though non-standard, is nonetheless assured
by the fluctuation-dissipation theorem, which relates linear
response functions to the frequency spectrum of intrinsic
fluctuations \cite{Kubo}. In a typical SNS experiment, random spin
fluctuations $\delta S_z(t)$ in an equilibrium sample impart Faraday
rotation (optical polarization rotation) fluctuations $\delta
\theta(t)$ on a probe laser, which can be measured with high
sensitivity. In the frequency domain, the peak
positions, widths, and amplitudes of this Faraday rotation (FR)
noise reveal the detailed dynamical properties of the spins such as
$g$-factors, coherence times, and relaxation rates -- without (in
principle) ever exciting or pumping the spin system itself. This
latter appealing aspect arises because spin detection via FR depends on the
system's dispersive indices of refraction (rather than absorption),
and in many systems it can form the basis for continuous quantum
nondemolition measurement \cite{Kuzmich, Takahashi}. Optical SNS has
been applied to alkali vapors \cite{CrookerGas}, electrons in bulk
\emph{n}-type GaAs \cite{Oestreich, CrookerGaAs}, quantum well
microcavities \cite{Poltavtsev}, and recently to electron and
hole spins in (In,Ga)As QD ensembles \cite{CrookerQD, Li}.

In all SNS studies reported to date, a single probe laser was used
to detect the intrinsic spin fluctuations of the system. For the
studies of QD ensembles \cite{CrookerQD, Li}, this probe laser was
tuned in wavelength to lie directly within the
inhomogeneously-broadened absorption/photoluminescence band of the
ensemble. Because individual QDs within the ensemble have very
narrow homogeneous linewidths $\gamma_h$ at low temperature
\cite{Gammon, Atature, Gerardot, WarburtonReview, Vamivakas}, the probe laser
is particularly sensitive to spin fluctuations from those
singly-charged QDs having charged exciton (trion) optical transition
energies close to the laser energy.

Consider, as depicted in Fig. 1a, an inhomogeneously-broadened QD
ensemble and two singly-charged dots (QDA and QDB) with different
trion transition energies within this ensemble. Per the usual
optical selection rules for spins in III-V or II-VI semiconductors
\cite{WarburtonReview, SelectionRules}, individual QDs within this
ensemble exhibit spin-dependent absorption spectra for right- and
left-circularly polarized light ($\alpha^{R,L}$, assumed to be
Lorentzian) that depend on the orientation of the resident spin. For
example,
\begin{equation}\label{eq:1}
\alpha^{R}(\omega,\omega_k)\propto\frac{\gamma_h/2}{(\omega-\omega_k)^2+(\gamma_h/2)^2}~,~\alpha^{L}=0
\end{equation}
if the resident spin has projection ``spin-up" along the probe laser
($S_z \| \vec{k}$), while $\alpha^R$ and $\alpha^L$ are swapped for
opposite spin projection. Here, $\omega$ is the photon energy,
$\omega_k$ is the energy of the charged exciton transition of the QD
in question, and $\gamma_h$ is its homogeneous absorption linewidth.
The Faraday rotation, $\theta$, that is imparted to a probe laser by
this spin-dependent optical transition scales as the difference
between the associated indices of refraction, $n^R$ and $n^L$:
\begin{equation}\label{eq:2}
\theta(\omega,\omega_k)\propto n^R - n^L \propto \pm \frac{\omega-\omega_k}{(\omega-\omega_k)^2+(\gamma_h/2)^2}.
\end{equation}
Therefore when the resident spin in a QD fluctuates randomly in
time, $\theta(t)$ also fluctuates. In thermal equilibrium and in
zero magnetic field its time average is of course zero
($\langle\theta(t)\rangle=0$), but its \emph{variance}
$\langle\theta^2(t)\rangle$ -- the FR noise power -- is nonzero and
is peaked at photon energies $\pm \gamma_h/2$ away from the QD
resonance, and decays as $|\omega-\omega_k|^{-2}$ for large
detuning. (Note also that spin fluctuations induce no FR
noise exactly on resonance when $\omega=\omega_k$, because $\theta$
is an odd function). Therefore the probe laser depicted in Fig. 1a,
which has photon energy close to the QDA trion resonance, is much
more sensitive to spin (FR) fluctuations from QDA than from QDB.

We exploit this spectral selectivity, and also the fact that
fluctuations of spins in different dots are nominally
\emph{uncorrelated} in time, to directly obtain $\gamma_h$ in an
ensemble measurement, using a low-intensity (linear) optical
experiment based on spin noise. Specifically, we use \emph{two}
different probe lasers (``1" and ``2") that are tuned within the
broad absorption band of the QD ensemble, and measure the degree of
correlation between the FR fluctuations that are imparted on the two
lasers [$\theta_1(t)$ and $\theta_{2}(t)$]. The two co-propagating
probe lasers are incident on the same photodetectors, and the total
measured FR is just the sum $\theta_1(t) +\theta_2(t)$. The variance of
the FR noise (\emph{i.e.}, the measured noise power) is therefore
\begin{equation}\label{eq:variance}
\langle[\theta_1(t) +\theta_2(t)]^2 \rangle =\langle\theta_1^2(t)
\rangle +\langle \theta_2^2(t)\rangle + 2\langle \theta_1(t)
\theta_2(t)\rangle.
\end{equation}
As depicted in Fig. 1b, if the detuning $\Delta \omega$ between the
two lasers is large ($\Delta\omega\gg\gamma_h$), then each laser
probes a different and independent subset of fluctuating spins, and
$\theta_1(t)$ and $\theta_{2}(t)$ are uncorrelated and add
incoherently (\emph{i.e.}, the cross-term in Eq. 3 averages to
zero). In contrast, if $\Delta\omega \leq \gamma_h$ (see Fig. 1c),
then the lasers measure predominantly the same QDs, $\theta_1(t)
\approx \theta_{2}(t)$ and the FR fluctuations are correlated,
giving larger measured noise power.

\begin{figure}
\centering
\includegraphics[width=0.45\textwidth]{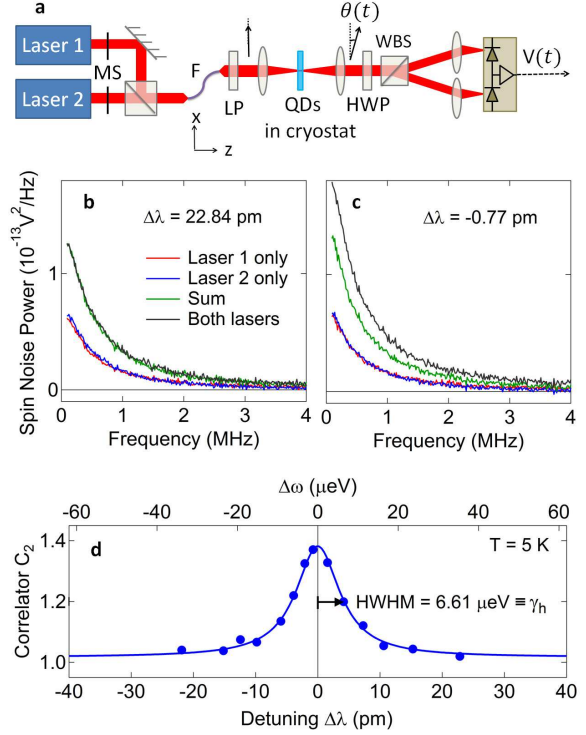}
\caption{\textbf{Measuring the two-color spin noise correlator
$C_2(\Delta \omega)$ and QD homogeneous linewidth $\gamma_h$}.
\textbf{(a)}, Experimental schematic: narrowband probe lasers 1 and 2
are combined in a single-mode polarization-maintaining fiber (F).
The probe light is focused through the ensemble of singly-charged
QDs, where hole spin fluctuations impart FR noise on the transmitted
light, which is measured by balanced photodiodes. Here, MS are
mechanical shutters, HWP is a half-wave plate and WBS is a Wollaston
beam splitter. HWP can be replaced by a quarter-wave plate to
measure ellipticity noise. \textbf{(b,c)}, Raw spin noise power
spectra for the case of large and small laser detuning
($\Delta\omega > \gamma_h$ and $\Delta\omega < \gamma_h$), giving
uncorrelated and correlated spin noise, respectively. $T$=5~K. The
red and blue spectra were detected using \emph{only} laser 1 or
laser 2, respectively. The green trace is their mathematical sum.
The black trace was measured with \emph{both} lasers on
simultaneously. \textbf{(d)}, Two-color spin noise correlator $C_2$ as
a function of the detuning between the probe lasers. The solid line
is a Lorentzian fit, and its HWHM reveals the underlying homogeneous
linewidth $\gamma_h$ of the QDs.}\label{fig:2}
\end{figure}

The two-color spin noise experiment is depicted in Fig. 2a. The
low-power outputs from two tunable continuous-wave probe lasers (1
and 2, each having $<$10~MHz linewidth) are combined and launched
through a single-mode polarization-maintaining fiber to ensure spatial overlap. The light is linearly polarized and focused weakly
through an ensemble of singly-charged QDs (typical spot sizes are
$\sim$25 $\mu$m, with $\sim$100 $\mu$W power from each laser).
Stochastic fluctuations of the hole spins in the QDs generate FR
fluctuations $\theta(t)=\theta_1(t) + \theta_2(t)$ on the
transmitted probe beam, which are measured with balanced
photodiodes. The output voltage $V(t) \propto \theta(t)$ is
continuously digitized and Fourier-transformed in real time
\cite{CrookerQD} to obtain the noise power density spectrum (shown
here in units of $V^2$/Hz). Mechanical shutters control whether the
probe beam is composed of laser 1, 2, or both. The samples are
lightly $p$-type (In,Ga)As/GaAs QDs grown by molecular beam epitaxy
(see Methods). Owing to statistical variations in QD size and
composition, the ensemble PL spectrum is strongly inhomogeneously
broadened ($\gamma_{inh} \sim 20$~meV), and is peaked at $\sim$1.385
eV (895 nm).

Figure 2b shows the measured power spectra of hole spin noise for
the case of large detuning between probe lasers ($\Delta
\lambda$=22.84~pm, or $\Delta \omega$=35.36~$\mu$eV). The red and
blue noise spectra were acquired using the individual lasers 1 and 2
alone. The noise spectra are Lorentzian with $\sim$500 kHz
half-width, indicating long hole spin relaxation times of $\sim$300
ns, in agreement with previous (single-probe) noise studies of
similar QD ensembles \cite{Li}. The green spectrum is just the
mathematical sum of these two single-probe measurements. The black
spectrum is the spin noise acquired using \emph{both} lasers 1 and 2
simultaneously. Here, this black trace overlaps almost exactly with
the green, indicating that in this case the FR fluctuations on the
two lasers are \emph{uncorrelated}: the noise power with both lasers
is simply the sum of the noise power from the two individual probe
lasers because the interference term in Eq. 3 vanishes.

In marked contrast, Fig. 2c shows the case for small detuning
between the probe lasers ($\Delta \lambda$=-0.77~pm, or $\Delta
\omega$=-1.19~$\mu$eV).  Here, the spin noise power measured with
both lasers simultaneously is \emph{greater} than the sum of the
noise power measured by lasers 1 and 2 individually, indicating that
the FR noise encoded on the two probe lasers is at least partially
correlated such that the interference term in Eq. 3, $\langle
\theta_1(t) \theta_2(t)\rangle$, exceeds zero.

We define the two-color spin correlator as
\begin{equation}\label{eq:3}
C_2(\omega_1,\Delta\omega)\equiv\frac{P_{\hbox{both}}(\omega_1,\Delta\omega)}{P_1(\omega_1)+P_2(\omega_1+\Delta\omega)},
\end{equation}
expressed here as a function of the photon energy $\omega_1$ of
probe laser 1 and the detuning $\Delta\omega$ between lasers 1 and
2. Here, $P_{1}(\omega_1)$ and $P_2(\omega_2=\omega_1 +
\Delta\omega)$ are the total spin noise power measured by individual
probe lasers 1 and 2 respectively (computed via the area under the
measured noise spectra in, \emph{e.g.}, Figs. 2b,c).
$P_{\hbox{both}}(\omega_1,\Delta\omega)$ is the total spin noise
power measured using both lasers simultaneously. Following Eq. 3, we
can therefore expect the two-color spin correlator $C_2$=1 if the
noise on the two lasers is uncorrelated, but increases to 2 when the
noise is perfectly correlated and $\theta_1(t)=\theta_2(t)$.

Figure 2d shows the measured $C_2$ versus detuning $\Delta \omega$.
It is clearly peaked at $\Delta \omega$=0 as expected, and falls
rapidly to unity as $|\Delta \omega|$ increases. It fits very well
to a Lorentzian function with a very narrow half-width at
half-maximum (HWHM) of 4.27~pm, or 6.6~$\mu$eV. As shown immediately
below, \emph{the key point of two-color spin noise spectroscopy} is
that the half-width of $C_2(\Delta\omega)$ directly reveals the
underlying homogeneous linewidth $\gamma_h$ of the singly-charged
QDs in the ensemble.

It is straightforward to show that $C_2(\Delta \omega)$ is expected
to exhibit a Lorentzian shape with HWHM equal to $\gamma_h$. First,
note that the total FR noise power detected by a single probe laser
at energy $\omega_i$ is given by integrating up the FR noise power
$\langle \theta^2(t) \rangle$ generated by all the QDs (at energies
$\omega_k$) in the ensemble:

\begin{equation}\label{eq:4}
P_i(\omega_i)=\int_{0}^{\infty}\theta_i^2(\omega_i,\omega_k)\rho(\omega_k)d\omega_k,~~i=1,2
\end{equation}
where $\theta_i(\omega_i, \omega_k)$ has the same form as in Eq. 2,
and $\rho(\omega_k)$ is the inhomogeneously-broadened distribution
of QD energies characterized by $\gamma_{inh}$. Effectively, this
corresponds to integrating the square of the dotted red or blue
curves in Figure 1b, weighted by $\rho(\omega_k)$. As discussed
earlier, most of the noise power comes from those QDs with
resonances close to the probe laser.

When \emph{both} lasers probe the QDs simultaneously (at $\omega_1$
and $\omega_2 = \omega_1 + \Delta \omega$), the total spin noise
power is
\begin{equation}\label{eq:6}
\begin{split}
&P_{\hbox{both}}(\omega_1,\Delta\omega)\\
=&\int_{0}^{\infty}[\theta_1(\omega_1,\omega_k)+\theta_2(\omega_1+\Delta\omega, \omega_k)]^2
\rho(\omega_k)d\omega_k\\
=&P_1(\omega_1)+P_2(\omega_1 + \Delta\omega)\\
&+2\int_{0}^{\infty}\theta_1(\omega_1,\omega_k)\theta_2(\omega_1+\Delta\omega, \omega_k)
\rho(\omega_k)d\omega_k,
\end{split}
\end{equation}
which corresponds to integrating the square of the \emph{sum} of the
dotted red and blue lines in Figs. 1b,c. With both lasers tuned
within the inhomogeneously-broadened ensemble, and since
$\gamma_{inh}$ greatly exceeds both $\gamma_{h}$ and typical
detunings $\Delta \omega$, $\rho(\omega_k)$ can be approximated by a
uniform distribution [$\rho(\omega_k) \sim 1$] and $C_2$ can be
calculated analytically:
\begin{equation}\label{eq:7}
C_2(\Delta\omega)=1+\frac{\gamma_h^2}{\Delta\omega^2+\gamma_h^2},
\end{equation}
which is a Lorentzian with HWHM=$\gamma_h$. (We note that if the
lasers are tuned far \emph{outside} of the inhomogeneous absorption
band then $C_2(\Delta \omega)$ is expected to be constant, a
scenario that will be discussed and modeled later).

Thus, using only low-power continuous-wave probe lasers and
performing passive measurements, the homogeneous linewidth
$\gamma_h$ of singly-charged QDs is revealed in an ensemble
measurement. Crucially, this is made possible because we measure
fluctuations and correlations, rather than conventional
time-averaged linear responses. The measured value ($\gamma_h = 6.6$
$\mu$eV) is in rather good agreement with recent nonlinear four-wave
mixing studies \cite{MoodyTrion} of very similar positively-charged (In,Ga)As QD
ensembles ($\gamma_h = 8 \pm 2$ $\mu$eV at 10~K) and also
with absorption measurements of individual positively-charged InGaAs QDs ($\gamma_h \sim 5$ $\mu$eV at 4.2~K in ref. \cite{Gerardot,
WarburtonReview}, and $\gamma_h \sim 2$~GHz $\equiv 8.2$ $\mu$eV at 5~K in ref. \cite{GreilichNatPhys}). It is less than $\gamma_h$ inferred from coherent control studies of hole qubits in single InAs-based QDs (6.7~GHz $\equiv$27 $\mu$eV at 1.6~K) \cite{DeGreve}, but in all cases the different QD growth and device fabrication conditions make direct comparisons difficult.

\begin{figure}
\centering
\includegraphics[width=0.38\textwidth]{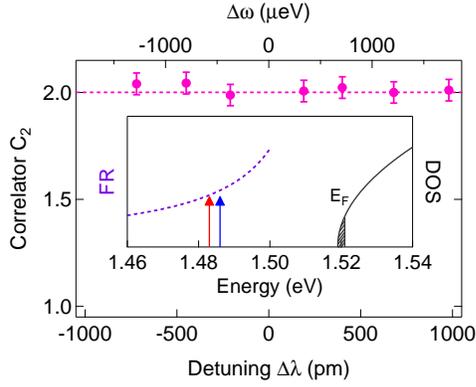}
\caption{\textbf{Control experiment on $n$-type bulk GaAs}
($n=3.7\times10^{16}$ cm$^{-3}$). $T$=10~K. Here the spin noise of
electrons at the bottom of the conduction band can be probed at
photon energies well below the GaAs bandgap ($\omega_1 \sim
1.485$~eV). The correlator $C_2$ is always 2, independent of the
detuning $\Delta \omega$. The inset depicts the density of states
(DOS) in GaAs, the Fermi level of the electron sea ($E_F$), and the
FR that is induced at sub-gap energies when these electrons are
polarized (dashed line).}\label{fig:3}
\end{figure}

We note that both the principle and the technique of two-color spin
noise spectroscopy -- as well as the information obtained -- are essentially
different than the methods for `optical spectroscopy of spin noise'
discussed recently by Zapasskii \emph{et al} \cite{Zapasskii}. Ref.
\cite{Zapasskii} describes how single-probe SNS measurements can be used, \emph{e.g.}, to tell the difference
between homogeneous and inhomogeneously-broadened lines, or to infer
whether the ratio $\gamma_{inh}/\gamma_h$ is changing in response to
some external parameter like temperature. However, ref.
\cite{Zapasskii} concerns exclusively single-probe SNS experiments
(not multi-probe), and measurements of noise power only (not
correlations). Most importantly, direct measurements of $\gamma_h$
are not possible, in contrast to the case here.

To validate this two-color noise technique, Fig. 3 shows a control
experiment on bulk \emph{n}-type GaAs. As shown in earlier studies
\cite{CrookerGaAs, Glasenapp}, spin fluctuations of electrons in the
conduction band of $n$-GaAs generate FR fluctuations that can be
detected at photon energies far below the low temperature GaAs band-gap
(\emph{e.g.}, at sub-gap wavelengths from 830-900~nm), owing to the
long tails of the dispersive indices of refraction (see inset).
Importantly, FR noise detected at these low energies derives
primarily from the \emph{same} fluctuating electron spins in the
conduction band, independent of probe laser wavelength. Thus, we
expect correlated noise and $C_2(\Delta \omega)$=2 in a two-color
noise experiment on \emph{n}-GaAs, independent of $\Delta \omega$.
Exactly this behavior was observed and confirmed, as shown in Fig.
3. Note that this experiment corresponds to a situation where both
probe lasers are tuned well outside of any inhomogeneously-broadened
absorption band, a situation modeled later in Fig. 6.

\begin{figure}
\centering
\includegraphics[width=0.49\textwidth]{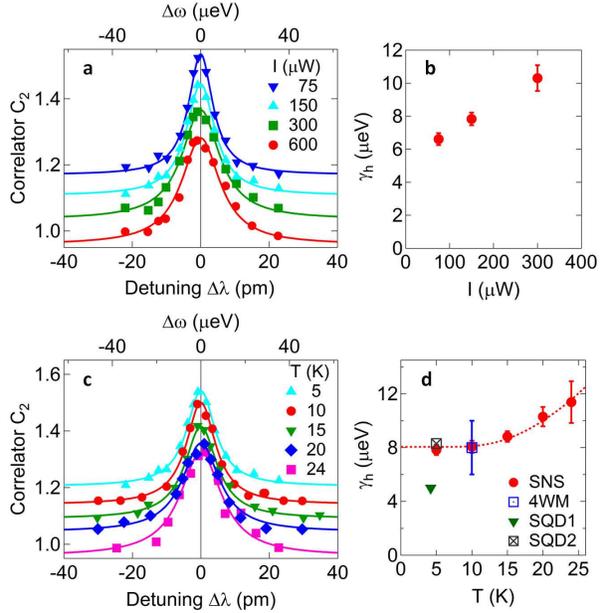}
\caption{\textbf{Measuring the two-color spin noise correlator
$C_2(\Delta \omega)$ and homogeneous linewidth $\gamma_h$ versus
laser intensity and temperature.} \textbf{(a)}, $C_2(\Delta \omega)$
at different probe laser intensities $I$ at 5 K (vertically offset
for clarity). \textbf{(b)}, The corresponding homogeneous linewidths
of the QDs. \textbf{(c)}, $C_2(\Delta\omega)$ at various temperatures
(vertically offset). $I=$150 $\mu$W. \textbf{(d)}, The extracted
homogeneous linewidth $\gamma_h$ (red dots). The dotted line is a
fit to the data following the model in \cite{Borri}. Also shown is
$\gamma_h$ determined by non-linear four-wave mixing measurements of
positively-charged exciton transitions (blue square) on similar
(In,Ga)As QD ensembles at 10 K \cite{MoodyTrion}, and from absorption studies \cite{Gerardot, GreilichNatPhys} of hole-doped single QDs (triangle, black square).}\label{fig:4}
\end{figure}

Although Fig. 2d shows that $C_2(\Delta\omega)$ for the QD ensemble
exhibits the anticipated Lorentzian lineshape and reveals the
expected $\gamma_h$, we note that its peak value is only $\sim$1.4,
which is less than the value of 2 that was observed in the $n$-GaAs
control sample and which is expected from Eq. (\ref{eq:7}). This is
because the probe lasers are tuned directly within the absorption
band of the QD ensemble, and are therefore unavoidably pumping those
QDs that are resonant with the lasers. In this regime the probe
lasers cannot be considered completely non-perturbing, as evidenced
by the fact that the measured spin noise density (in units of
$V/\sqrt{\textrm{Hz}}$) increases only sub-linearly with laser
intensity $I$ (and the spin noise \emph{power} density in units of
$V^2$/Hz increases less-than-quadratically with $I$). For the very
narrowband lasers used in these studies ($<$10 MHz), independent
experiments confirm that the measured noise power increases as
$\sim$$I^{1.5}$ instead of $I^2$. This is consistent with the peak
value of the correlator $C_2(\Delta \omega \sim 0)$ measured in Fig.
2d ($2^{1.5}/2 \approx 1.4$).

It is well established from optical studies of single QDs and also
from nonlinear optical studies of QD ensembles that the measured
$\gamma_h$ is strongly dependent on the intensity of the probing
light \cite{Vamivakas, Berry}.  Figures 4a,b explore the influence
of probe laser intensity $I$ on the two-color noise correlator
$C_2(\Delta \omega)$ at 5~K. The inferred $\gamma_h$ increases with
$I$, similar to past studies \cite{Vamivakas, Berry}, likely due to
the resonant QD pumping effects and excitation-induced broadening
discussed above. In the limit of zero laser intensity, these data
suggest $\gamma_h \sim$ 6 $\mu$eV for the positively-charged trion
transition in these QDs, which is consistent with prior results
\cite{Gerardot, WarburtonReview, MoodyTrion, GreilichNatPhys} as
discussed above.

It is also well known that $\gamma_h$ broadens with increasing
temperature in epitaxial QDs due to interactions with phonons.
Figures 4c,d shows the temperature dependence of $C_2(\Delta
\omega)$ and $\gamma_h$. $\gamma_h$ is nearly constant below 10~K,
but increases at higher temperatures. This overall trend agrees very
well with available reported data on the temperature dependence of
exciton linewidths in undoped but otherwise very similar (In,Ga)As
QDs \cite{Borri, BorriReview}, and also on interfacial GaAs QDs
\cite{Moody}, both of which were measured by nonlinear ultrafast
four-wave mixing techniques. For a direct visual comparison Fig. 4d also
plots $\gamma_h$ determined from other studies of positively-charged (In,Ga)As QDs discussed above \cite{Gerardot, WarburtonReview, MoodyTrion, GreilichNatPhys}, with which our results are also in quite reasonable agreement, confirming the viability and utility of this low-power noise-based optical technique.

\begin{figure}
\centering
\includegraphics[width=0.49\textwidth]{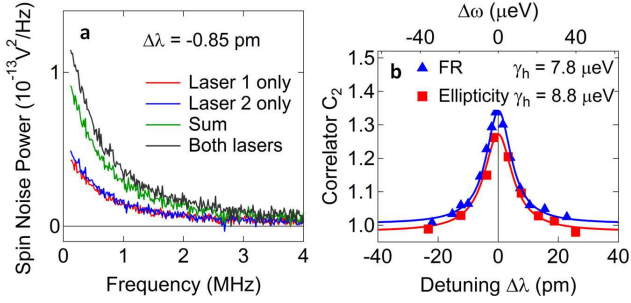}
\caption{\textbf{Spin noise can also be measured via ellipticity
fluctuations} of the probe laser(s). \textbf{(a)}, Raw ellipticity
noise power spectra at 5 K, $I=$150 $\mu$W. \textbf{(b)}, Directly
comparing $C_2(\Delta \omega)$ from FR and ellipticity spin noise
measurements, using the same probe laser power.}\label{fig:5}
\end{figure}

We also show that measurements of $\gamma_h$ are possible by
measuring \emph{ellipticity} fluctuations imparted on the transmitted probe light (instead of FR
fluctuations). Whereas spin
fluctuations induce FR noise via the dispersive real part of the QD
dielectric function (\emph{i.e.}, the indices of refraction
$n^{R,L}$), ellipticity noise is associated with fluctuations of the
imaginary part (\emph{i.e.}, the absorption, $\alpha^{R,L}$).
Ellipticity and FR noise are linked via Kramers-Kronig relations and
are therefore related. By replacing the half-wave plate in Figure 2a
with a quarter-wave plate, ellipticity noise is measured. Figure 5
shows raw ellipticity noise power from these QDs at 5~K, along with
$C_2(\Delta \omega)$.  Following a similar analysis, the ellipticity
correlator is found to have the same functional form as in Eq. 7,
\emph{i.e.}, a Lorentzian line shape with HWHM=$\gamma_h$. For a
given laser intensity $I$, we find that $\gamma_h$ obtained via ellipticity noise
is slightly larger than $\gamma_h$ determined from FR noise (Fig.
5b), a likely consequence of the excitation-induced broadening
discussed above and the fact that ellipticity is by definition more
sensitive to absorption and therefore to QDs exactly on resonance
with the probe lasers.

\begin{figure}
\centering
\includegraphics[width=0.45\textwidth]{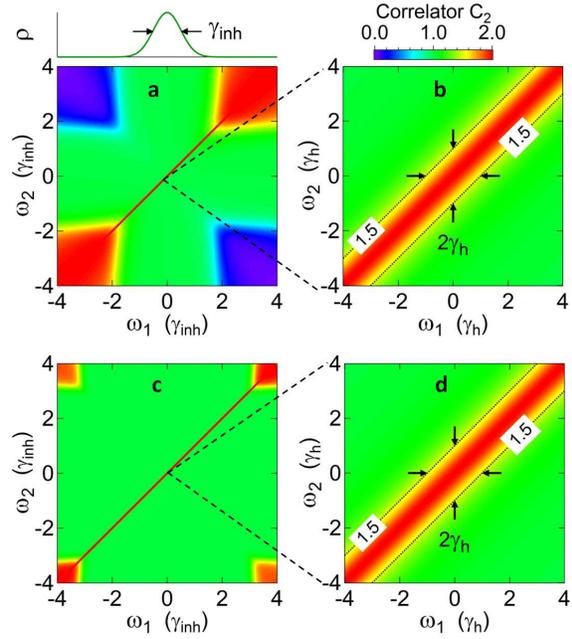}
\caption{\textbf{Modeling the two-color spin noise correlator $C_2$}
as a function of the two probe laser energies $\omega_1$ and
$\omega_2$ (horizontal and vertical axis, respectively). We consider
an inhomogeneously-broadened Gaussian band with $\gamma_{inh} =
100\gamma_h$. \textbf{(a)}, The case when FR noise is detected; see
text for details. \textbf{(b)}, An expanded view along the diagonal
where $\omega_1 \sim \omega_2$ (note change in scale). \textbf{c,d},
The case when ellipticity noise is detected. }\label{fig:6}
\end{figure}

Finally, Fig. 6 shows model calculations of the two-color noise
correlator $C_2$ over a broad range of the two probe laser energies
$\omega_1$ and $\omega_2$, using Eqs. 4-6.  For simplicity we
consider an idealized inhomogeneously-broadened Gaussian absorption band with
$\gamma_{inh}=100\gamma_h$. In the usual case where the probe lasers
detect spin fluctuations via Faraday rotation, Fig. 6a shows that
when $\omega_1$ and/or $\omega_2$ is tuned within this broad
inhomogeneous band, the spin noise is everywhere uncorrelated
($C_2$=1) except along the thin diagonal line corresponding to
$\omega_1 \sim \omega_2$ (more specifically, when $\Delta \omega
\lesssim \gamma_h$). Expanding this region (Fig. 6b), this thin band
has half-width equal to $\gamma_h$, as expected from Eq. 7 and as
experimentally observed in Figures 2 and 4. Interestingly, however,
when both $\omega_1$ and $\omega_2$ lie well \emph{outside} this
broad absorption band, then neither probe laser is preferentially
sensitive to nearly-resonant QDs (because there are no QDs at the
probe energies). In this case, both lasers are sensitive to the
total FR fluctuations from the entire ensemble, and their
fluctuations will be correlated ($C_2=2$) \emph{independent of}
$\Delta\omega$ (or anti-correlated giving $C_2=0$ if the lasers are
on opposite `sides' of this idealized band, because $\theta(\omega)$ is an odd
function). This was the precisely the case for the two-color spin
noise measurement on the $n$-type GaAs control sample shown in Figure
3. A similar modeling can be performed for the case when spin
fluctuations are measured via ellipticity noise, as shown in Figures
6c,d and as experimentally observed in Figure 5. (Note, however, that
although $C_2$ can be calculated for all $(\omega_1, \omega_2)$, ellipticity noise must generally be measured
within the absorption band because otherwise the noise signals
themselves become extremely small.)

In summary, we have introduced and demonstrated the new optical
technique of two-color spin noise spectroscopy. Despite being a
low-power optical method that does not rely on excitation, pumping,
or perturbation of the ensemble, it can directly reveal,\emph{
e.g.}, the underlying homogeneous linewidths of QDs that are
otherwise obscured in strongly inhomogeneously-broadened optical
spectra. This is because these methods are based on
\emph{fluctuations and correlations}, rather than on measurement of
the standard time-averaged response functions that are measured by
conventional linear optical spectroscopy. Though demonstrated here
for the specific case of semiconductor QDs and spin fluctuations,
the general principle should be broadly applicable to many classes
of systems and materials in which intrinsic fluctuations can be
detected -- and may prove especially attractive for those systems where the individual constituents
cannot easily be isolated, or for which intense non-linear
perturbation is undesired. Moreover, these noise-based techniques
can equally well be applied to -- and have great promise for --
studies of correlations in \emph{interacting} systems, such as
coupled quantum-dot or nanocrystal systems, spin-exchange
interactions in atomic gases, or spin-spin interactions in chemical
and biological systems. A further interesting extension of these
ideas is to measure not only temporal but also \emph{spatial}
correlations of systems exhibiting collective excitation (\emph{e.g.}
magnon, plasmon, or polariton systems), by two or more probes
with tunable spatial separation \cite{Pershin}.

\section{Methods}
\textbf{Quantum dot samples.} Self-assembled InAs/GaAs QDs were grown by molecular
beam epitaxy on (001) GaAs substrates, and then thermally annealed
at 940 $^\circ$C for 30 seconds. Annealing interdiffuses indium and
gallium and thus increases the size of the QDs and decreases the
depth of the confining potential, blue-shifting the QD
absorption band to $\sim$900~nm. The sample contains 20 layers of
QDs, separated by 60 nm GaAs barriers, with QD density of $10^{10}$
cm$^{-2}$ in each layer. The sample is weakly \emph{p}-type due to
background carbon doping; we estimate that $\sim$10\% of the QDs
contain a single resident hole. The measured spin noise arises from
stochastic fluctuations of the resident holes trapped in the
singly-charged subset of QDs. The control experiment was carried out
on a bulk \emph{n}-GaAs wafer with electron density
$3.7\times10^{16}$ cm$^{-3}$.

\textbf{Experimental setup.} The samples are mounted on the cold
finger of a liquid helium optical cryostat. Two tunable narrowband
continuous-wave diode lasers (from Toptica and New
Focus, both with $<$10~MHz linewidth) are tuned in wavelength near
the center of the inhomogeneously-broadened PL/absorption spectrum
of the QD ensemble. In all the data shown, the first laser remains
at a fixed frequency of $\omega_1=1.387$ eV (896.3 nm); the second
one is detuned from the first one by $\Delta\omega$. $\gamma_h$ did not vary appreciably for different values
of $\omega_1$. For the control experiment on bulk \emph{n}-GaAs, we
use two continuous-wave Ti:sapphire lasers (from Coherent). The
lasers were tuned $\sim$30~meV below the low-temperature absorption
edge of bulk GaAs ($E_{gap}=1.515$ eV or 818 nm). The spin noise
signal is continuously digitized and processed by an FPGA \cite{CrookerQD} to obtain the spin
noise power spectrum.


\subsection{Acknowledgements}
We are particularly indebted to M. Boshier for the loan of a tunable
laser system. We also thank N. Sinitsyn and D.L. Smith for helpful
discussions, and gratefully acknowledge D. Reuter and A. D. Wieck
for the QD sample. This work was supported by the Los Alamos
LDRD program (under the auspices of the US DOE, Office of Basic
Energy Sciences, Division of Materials Sciences and Engineering),
and also by the Deutsche Forschungsgemeinschaft.

\end{document}